\documentclass[12pt]{article}
\usepackage{hyperref}
\usepackage{amsmath,amssymb}
\usepackage{graphicx}
\usepackage{color}
\definecolor{myblue}{rgb}{0.14,0.11,0.49}
\definecolor{myred}{rgb}{0.74,0.22,0.15}
\definecolor{mygreen}{rgb}{0.05,0.52,0.42}
\definecolor{myyellow}{rgb}{0.96,0.92,0.13}
\definecolor{myorange}{rgb}{1,0.61,0.36}
\definecolor{mypurple}{rgb}{0.71,0.02,1}
\definecolor{noir}{gray}{0.} % black

\definecolor{htc}{rgb}{1,1,1} % heading text colour
\newcommand{\Mat}[1]{{{\boldsymbol{#1}}}}

\def\be{\begin{equation}}
\def\ee{\end{equation}}
\def\bea{\begin{eqnarray}}
\def\eea{\end{eqnarray}}
\def\bc{\begin{center}}
\def\ec{\end{center}}
\def\bi{\begin{itemize}}
\def\ei{\end{itemize}}
\def\bs{\begin{slide}}
\def\es{\end{slide}}

\def\iC{\mathrm{i}}
\def\noi{\noindent}
\title{Lorentz-invariant second-order tensors and an irreducible set of matrices}
\author{
Mayeul Arminjon\\
\small\it Univ. Grenoble Alpes, CNRS, Grenoble INP%\footnote{\ Institute of Engineering Univ. Grenoble Alpes}
, 3SR, F-38000 Grenoble, France
} 
\date{}
			     
\begin{document}
%%%%%%%%%%%%%%%%%%%%%%%%%%%%%%%%%%%%%%%%%%%%%%%%%%%%%%%%%%%%%%%%%%%%%%%%%%%%%%%%

\maketitle

\begin{abstract}
\noi We prove that, up to multiplication by a scalar, the Minkowski metric tensor is the only second-order tensor that is Lorentz-invariant. To prove this, we show that a specific set of three $4\times 4$ matrices, made of two rotation matrices plus a Lorentz boost, is irreducible.
\end{abstract}

%%%%%%%%%%%%%%%%%%%%%%%%%%%%%%%%%%%%%%%%%%%%%%%%%%%%%%%%%%%%%%%%%%%%%%%%%%%%%%%%
\section{Introduction}\label{Intro}
%%%%%%%%%%%%%%%%%%%%%%%%%%%%%%%%%%%%%%%%%%%%%%%%%%%%%%%%%%%%%%%%%%%%%%%%%%%%%%%%

It is a basic result of special relativity that the Minkowski metric tensor is invariant under the Lorentz group. The main aim of this paper is to prove that, up to a scalar, this property characterizes the Minkowski metric:

\paragraph{Theorem.}\label{Theorem} {\it Let $({\bf M},\Mat{\gamma }^0)$ be the four-dimensional Minkowski spacetime. Any $(0\quad 2)$ tensor on ${\bf M}$ that is invariant under the Lorentz group is a scalar multiple of the Minkowski metric tensor $\Mat{\gamma }^0$.}
\\

\noi (See Note \ref{LorentzianST} for the extension to a Lorentzian spacetime.) This result is not very surprising and seems to be heuristically known. For instance, after having introduced the classical totally antisymmetric fourth-order tensor, Maggiore \cite{Maggiore2005} states: ``The only other invariant tensor of the Lorentz group is $\eta _{\mu \nu }$ [the Minkowski metric, M.A.]; its invariance follows from the defining property of the Lorentz group, eq. (2.13)." (The latter equation is equivalent to Eq. (\ref{Lor-inv eta}) below.) Nevertheless, we saw neither a precise statement of the above Theorem nor a correct proof in the literature that we could find. The proof that we present here appeals to Schur's lemma (Section \ref{ProofTheorem}). However, to identify a relevant irreducible set of $4\times 4$ matrices in order to use Schur's lemma was not completely obvious. To prove the irreducibility of that set $\mathrm{S}$, we had to study in detail which are the invariant subspaces of each of the matrices that constitute $\mathrm{S}$ (Sect. \ref{ProofProp2}): although it is often easy to check that some subspace is invariant under some endomorphism (or some matrix), it is less trivial to identify the complete list of the invariant subspaces. To do that in the case at hand, we state and prove a result (\hyperref[Lemma 2]{Lemma 2}) about the invariant subspaces of a specific class of endomorphisms (Sect. \ref{ProofLemma2}).

%%%%%%%%%%%%%%%%%%%%%%%%%%%%%%%%%%%%%%%%%%%%%%%%%%%%%%%%%%%%%%%%%%%%%%%%%%%%%%%%
\section{Proof of the Theorem}\label{ProofTheorem}
%%%%%%%%%%%%%%%%%%%%%%%%%%%%%%%%%%%%%%%%%%%%%%%%%%%%%%%%%%%%%%%%%%%%%%%%%%%%%%%%

A $(0\ 2)$  second-order tensor $\Mat{T}$ at some point $X\in {\bf M}$
%\footnote{\ Formally: an element of the vector space $\mathrm{T}{\bf M}^\circ _X\otimes \mathrm{T}{\bf M}^\circ _X$, where $\mathrm{T}{\bf M}^\circ _X$ is the cotangent vector space at $X\in {\bf M}$.} 
 is Lorentz-invariant iff, in any Cartesian coordinates $x^\mu$ on ${\bf M}$, and for any $4\times 4$ real matrix $L=(L^\mu _{\ \,\nu})$ belonging to the (real) Lorentz group ${\sf O(1,3)}$, we have
 \footnote{\  \label{LorentzianST}
This definition and the Theorem extend immediately to any Lorentzian spacetime $(\mathrm{V},\Mat{\gamma })$, by considering, instead of Cartesian coordinates, coordinates that are Cartesian at the given point $X \in \mathrm{V}$, i.e., such that $\gamma _{\mu \nu }(X) = \eta _{\mu \nu }$.
}
\be
T'_{\mu \nu } := L^\rho  _{\ \,\mu}\,T_{\rho \sigma }\,L^\sigma  _{\ \,\nu} = T_{\mu \nu },
\ee
or ($T =(T_{\mu \nu })$ being the component matrix of $\Mat{T}$ at $X$ in the coordinates $x^\mu$)
\be\label{Lor-inv T}
L^T\,T\,L = T.
\ee
In particular, the Minkowski metric tensor $\Mat{\gamma }^0$, with component matrix $\eta :=\mathrm{diag}(1,-1,-1,-1)$ in any Cartesian coordinates, is of course a Lorentz-invariant $(0\ 2)$ second-order tensor on ${\bf M}$, since by definition a matrix $L$ belongs to the Lorentz group ${\sf O(1,3)}$ iff
\be\label{Lor-inv eta}
L^T\,\eta \,L = \eta .
\ee
Because the matrix $\eta $ is its own inverse: $\eta \,\eta ={\bf 1}_4:=\mathrm{diag}(1,1,1,1)$, we deduce from (\ref{Lor-inv eta}) that
\be\label{Lor-inv eta^-1}
L^{-1}\,\eta \,(L^T)^{-1} = \eta. 
\ee 
Multiplying on the right each side of (\ref{Lor-inv T}) by the corresponding side of (\ref{Lor-inv eta^-1}), we get:
\be
L^T\,T\,\eta \,(L^T)^{-1}  = T\,\eta ,
\ee
or
\be
L^T\,(T\,\eta)   = (T\,\eta )\,L^T.
\ee
That is, the matrix $M:=T\,\eta$ commutes with the transpose of any matrix $ L$ in the Lorentz group. The \hyperref[Theorem]{Theorem} results immediately from this and from the following two statements. \hfill $\square$\\

\noi {\bf Lemma 1 (Schur's lemma)} (e.g. \cite{Godement1966, Richtmyer2012}). {\it Let $M$ be a $k\times k$ complex matrix which commutes with any matrix in an irreducible set $\mathrm{S}$ of $k\times k$ matrices, i.e., in a set $\mathrm{S}$ of $k\times k$ complex matrices such that no nontrivial subspace of $\mathbb{C}^k$ is invariant under all mappings ${\bf x} \mapsto N\,{\bf x},\quad N\in \mathrm{S}$. Then $M$ is a complex multiple of the identity matrix ${\bf 1}_k$.}\\

\noi {\bf Proposition 1.} {\it The set of the matrices $L^T,\ L\in {\sf O(1,3)}$, is an irreducible set of $4\times 4$ complex matrices.}\\

\noi {\it Proof of Proposition 1.} In view of the sentence right after Eq. (\ref{Lz}) below, Proposition 1 is an immediate consequence of the following stronger result. \hfill $\square$\\

\paragraph{Proposition 2.}\label{Proposition 2} {\it Let the complex vector space $\mathbb{C}^4$ be endowed with its canonical basis $(e_\mu)\ (\mu=0,...,3)$, with $e_\mu =(\delta ^\nu _\mu)_{\nu =0,...,3}$, and identify an endomorphism of $\mathbb{C}^4$ with its matrix in that basis. The set $\mathrm{S}$ made by the two spatial rotations $L_i$ having axis $e_i$, $i=1,2$, each having a given angle $\theta_i$ with $0<\theta_i <\pi$, plus the Lorentz boost $L'_1$ in the direction $e_1$ with a given coefficient $\beta_v :=\frac{v}{c}$, $ 0<\beta_v <1$, is an irreducible set of three $4\times 4$ complex matrices.}\\

\section{Proof of Proposition 2}\label{ProofProp2}

The rotation matrix having axis $e_1$ and angle $\theta $ (with $\theta =\theta _1$ in the sequel) is 

\be\label{L_1}
L_1 = \begin{pmatrix} 
1 & 0  & 0 & 0\\
0 & 1  & 0 & 0\\
0 & 0 & \cos \theta & -\sin \theta \\
0 & 0 & \sin \theta & \cos \theta 
\end{pmatrix}.
\ee
The rotation matrices with axes $e_2$ and $e_3$ and angle $\theta $ are deduced from (\ref{L_1}) by the permutations $(1\ 2\ 3) \mapsto (2 \ 3\ 1)$ and $(1\ 2\ 3) \mapsto (3 \ 1\ 2)$ of the indices, respectively. The boost matrix in direction $e_1$ and with coefficient $\beta_v $ is
\be\label{L'_1}
L'_1 = \begin{pmatrix} 
\gamma_v  & -\gamma_v \beta_v   & 0 & 0\\
-\gamma_v \beta_v & \gamma_v   & 0 & 0\\
0 & 0 & 1 & 0 \\
0 & 0 & 0 & 1
\end{pmatrix}, \qquad \gamma_v := \frac{1}{\sqrt {1-\beta_v ^2}}.
\ee
Note that all matrices $L$ in the set $\{L_1,L_2,L_3,L'_1\}$ have real coefficients. Therefore, an endomorphism of either the real space $\mathbb{R}^4$ or the complex space $\mathbb{C}^4$ can be given by such a matrix $L$ in the same canonical basis $(e_\mu)$, which is also a basis of $\mathbb{R}^4$. The difference lies in the real or complex coefficients $z^\mu$ below:
\be\label{Lz}
z = z^\mu\, e_\mu \mapsto Lz = L^\mu_{\ \,\nu}\,z^\nu\, e_\mu.
\ee
Any matrix $L$ in the set $\{L_1,L_2,L_3,L'_1\}$ has the form $L'^T,\ L'\in {\sf O(1,3)}$, of course, because $L_i^T$ is the rotation matrix having axis $e_i$ and angle $-\theta_i$, which belongs to (the real group) ${\sf O(1,3)}$, and because $L'_1$ is symmetric and also belongs to ${\sf O(1,3)}$. \\

Henceforth, we consider only the complex vector space $\mathbb{C}^4$ and its complex subspaces, as well as their complex endomorphisms, including those defined by the matrices $L\in \mathrm{S}$. Clearly, the complex vector plane $\mathrm{Span}\{e_2,e_3\}$ or (in a shorter notation) $[e_2,e_3]$ is invariant under the rotation $L_1$. It is well known and easy to check that the restriction of  $L_1$ to $[e_2,e_3]$ has complex eigenvalues $\lambda _\pm=\exp\ (\pm \iC \theta_1 )$ (which here are distinct from one another and from $1$ since $0<\theta_1 <\pi$), with corresponding eigenvectors $a_\pm =e_2 \mp \iC e_3$. The complex eigenvalues of the rotation matrices with axes $e_2$ and $e_3$ are $\exp\ (\pm \iC \theta_i )\ (i=2,3)$, and the corresponding eigenvectors are deduced from $a_\pm$ by the permutations $(1\ 2\ 3) \mapsto (2 \ 3\ 1)$ and $(1\ 2\ 3) \mapsto (3 \ 1\ 2)$ of the indices, respectively. We will now show that the set $\mathrm{S}$ is an irreducible set of complex matrices, by using these facts about the eigenvectors of the $L_i$ 's, plus the lemma below --- whose proof is deferred to Sect. \ref{ProofLemma2} for convenience.

\paragraph{Lemma 2.}\label{Lemma 2} {\it Assume the finite-dimensional vector space $\mathrm{E}$ is the direct sum of two subspaces $\mathrm{F}$ and $\mathrm{G}$: $\mathrm{E}=\mathrm{F}\oplus \mathrm{G}$, each of which being invariant by the endomorphism $T$ of $\mathrm{E}$, with, moreover, $T_{\mid  \mathrm{F}}=\lambda \, \mathrm{Id}_\mathrm{F}$, and with $T_{\mid  \mathrm{G}}$ admitting a basis $(v_j)\ (j=1,...,n)$ of eigenvectors corresponding with pairwise distinct eigenvalues $\lambda _j$ such that, in addition, $\lambda _j \ne \lambda\ (j=1,...,n)$.\\

Then, each invariant subspace $\mathrm{W}$ of $\mathrm{E}$ by $T$ has the form
\be\label{u_i&v'_k}
\mathrm{W}= [(u_i)_{i\in \mathrm{I}};(v_j)_{j\in \mathrm{J}}]:=\mathrm{Span}\{(u_i)_{i\in \mathrm{I}};(v_j)_{j\in \mathrm{J}}\},
\ee
where $(u_i)_{i\in \mathrm{I}}$ ($0\leq p:=\mathrm{Card(I)}\leq \mathrm{dim\,F}$) is a family of linearly independent vectors of $\mathrm{F}$ ($p=0$ meaning that the family is empty), and where $(v_j)_{j\in \mathrm{J}}$ ($0\leq q:=\mathrm{Card(J)}\leq n=\mathrm{dim\,G}$) is a family of eigenvectors extracted from $(v_j)_{j=1,...,n}$ ($q=0$ meaning that the family is empty).}\\

Lemma 2 applies to the endomorphism $T=L_1$ of $\mathrm{E}:=\mathbb{C}^4$ given by the matrix (\ref{L_1}), with $\mathrm{F}:= [e_0,e_1]$ and $\mathrm{G}:=[e_2,e_3]$, the latter being stable by $L_1$ and admitting the basis of eigenvectors $(a_+,a_-)$. This allows us to easily write the complete list of the non-trivial vector subspaces of $\mathbb{C}^4$ which are invariant under the endomorphism $L_1$, Eq. (\ref{L_1}):

\bi
\item The (complex) ``lines" $[a]:=\mathbb{C}a$, where the vector $a\in \mathbb{C}^4$ is either:
\bi
\item a linear combination $a = \lambda e_0 + \mu e_1,\quad \lambda , \mu \in \mathbb{C}$, $\lambda \ne 0$ or $\mu \ne 0$;
\item or $a_+=e_2 - \iC e_3$ ;
\item or $a_-=e_2 + \iC e_3$.

\ei

\item The following ``planes" (2D complex subspaces):
\bi
\item (i) $[e_0,e_1]$;

\item (ii) $[e_2,e_3]= [e_2 - \iC e_3, e_2 + \iC e_3]$;

\item (iii) $[\lambda e_0 + \mu e_1, e_2 - \iC e_3]$, $\quad \lambda , \mu \in \mathbb{C}$, $\lambda \ne 0$ or $\mu \ne 0$;

\item (iv) $[\lambda e_0 + \mu e_1, e_2 + \iC e_3]$, $\quad \lambda , \mu \in \mathbb{C}$, $\lambda \ne 0$ or $\mu \ne 0$.
\ei

\item The following 3-spaces:

\bi

\item (a)  $[\lambda e_0 + \mu e_1, e_2, e_3]$, $\quad \lambda , \mu \in \mathbb{C}$, $\lambda \ne 0$ or $\mu \ne 0$;

\item (b) $[e_0,e_1,e_2 - \iC e_3]$;

\item (c) $[e_0,e_1,e_2 + \iC e_3]$.

\ei

\ei
The list of the non-trivial vector subspaces of $\mathbb{C}^4$ which are invariant under the endomorphism $L_i$ ($i=2$ or $i=3$) obtains by applying the permutation $(1\ 2\ 3) \mapsto (2 \ 3\ 1)$  or respectively $(1\ 2\ 3) \mapsto (3 \ 1\ 2)$ to the indices in the list above. Therefore, it is clear that the only line which is invariant under the rotations $L_1$ and $L_2$ (or under $L_1$ and $L_3$, or under $L_2$ and $L_3$) is $[e_0]$. However, that line is obviously not invariant under the boost (\ref{L'_1}). Thus no complex line is invariant under the set $\mathrm{S}$. As to the ``planes": it is clear also that none of the invariant planes by $L_1$ numbered (i) and (ii) in the list above is invariant under $L_2$ (nor by $L_3$, in fact). Still clear is the fact that an invariant plane by $L_1$, of the form (iii) or (iv): $[\lambda e_0 + \mu e_1, e_2 +\epsilon \iC e_3]$ ($\epsilon = \pm 1$), cannot coincide with either $[e_0,e_2]$ or $[e_3,e_1]$, which are invariant planes by $L_2$. The only remaining possibility to have an invariant plane by $L_1$ and by $L_2$ is if an invariant plane by $L_1$, of the form (iii) or (iv): $[\lambda e_0 + \mu e_1, e_2 +\epsilon \iC e_3]$ ($\epsilon = \pm 1$), can coincide with an invariant plane by $L_2$, of one of the corresponding forms: $[\lambda' e_0 + \mu' e_2, e_3 +\epsilon' \iC e_1]$ ($\epsilon' = \pm 1$, not necessarily $\epsilon =\epsilon '$). Thus the question is whether, for any $\alpha ,\beta \in \mathbb{C}$, one can find $\alpha' ,\beta' \in \mathbb{C}$ such that
\be
u(\alpha ,\beta ):= \alpha(\lambda e_0+\mu e_1) +\beta (e_2+\epsilon \iC e_3) = \alpha'(\lambda' e_0+\mu' e_2) +\beta' (e_3+\epsilon' \iC e_1),
\ee
that is, such that
\be\label{common plane}
\alpha \lambda =\alpha '\lambda ', \quad \alpha \mu = \iC \epsilon ' \beta ', \quad \beta = \alpha ' \mu ', \quad \iC \epsilon \beta = \beta '.
\ee
If $\epsilon =1$, we thus have $\beta '= \iC \beta $ by (\ref{common plane})$_4$, whence $\alpha \mu = -\epsilon ' \beta $ by (\ref{common plane})$_2$. Then if $\mu =0$, we must have $\beta =0$. If instead $\mu \ne 0$, we must have $\alpha =-\epsilon '\beta /\mu $. In either case, (\ref{common plane}) can apply only when either $\alpha $ or $\beta $ is determined by the other number, thus it cannot occur on the whole complex plane $[\lambda e_0 + \mu e_1, e_2 +\epsilon \iC e_3]$. The case $\epsilon =-1$ gives rise to the same discussion. Thus no 2-D subspace of $\mathbb{C}^4$ is invariant under both $L_1$ and $L_2$, {\it a fortiori} none is invariant under the set $\mathrm{S}$.\\

Let us finally look if there can be a 3-space invariant under the set $\mathrm{S}$, beginning with asking: which are, if any, the 3-spaces invariant under both $L_1$ and $L_2$? \\

1) We start the latter question by searching if a 3-space invariant under $L_1$, of the form (a) above, can coincide with a 3-space invariant under $L_2$, of the corresponding form. That is: can we have 
\be\label{3-space-0}
[\lambda e_0 + \mu e_1, e_2, e_3]= [\lambda' e_0 + \mu' e_2, e_3, e_1]? 
\ee
This is true iff, for any $\alpha ,\beta, \gamma  \in \mathbb{C}$, one can find $\alpha' ,\beta',\gamma ' \in \mathbb{C}$ such that
\be\label{3-space-1}
v:=\alpha(\lambda e_0+\mu e_1) +\beta e_2 + \gamma e_3 = \alpha'(\lambda' e_0+\mu' e_2) +\beta' e_3 + \gamma' e_1,
\ee
i.e.,
\be\label{3-space-2}
\alpha \lambda = \alpha ' \lambda ', \quad \alpha \mu =\gamma ', \quad \beta =\alpha ' \mu ', \quad \gamma =\beta '.
\ee
-- If $\lambda ' \ne 0$, we have from (\ref{3-space-2})$_1$: $\alpha ' = \alpha \lambda / \lambda '$. Then if $\mu ' \ne 0$, we get from (\ref{3-space-2})$_3$: $\alpha '=\beta /\mu '$, hence $\beta /\mu '= \alpha \lambda / \lambda '$. If instead $\mu ' = 0$, (\ref{3-space-2})$_3$ gives us $\beta =0$. In either case, $v$ in Eq. (\ref{3-space-1}) is assigned to depend at most on two parameters, hence (\ref{3-space-1}) cannot hold on the whole 3-space $[\lambda e_0 + \mu e_1, e_2, e_3]$.\\
\noi -- If $\lambda ' = 0$, we have from (\ref{3-space-2})$_1$: $ \alpha \lambda =0$. Then if $\lambda \ne 0$, this gives $\alpha =0$, so again (\ref{3-space-1}) cannot hold on the whole 3-space $[\lambda e_0 + \mu e_1, e_2, e_3]$ . If instead $\lambda =0$, then since we are considering the case $\lambda '=0$, necessarily $\mu \ne 0$ and $\mu ' \ne 0$ to have indeed a 3-space on both sides of (\ref{3-space-0}), thus this is the case that both of them coincide with the 3-space $[e_1, e_2, e_3]$, which is indeed invariant under $L_1$ and under $L_2$.\\%, and in fact also by $L_3$

2) Then we have to see if  a 3-space invariant under $L_1$, of the form (a) above, can coincide with a 3-space invariant under $L_2$ and corresponding with the cases (b) or (c), though of course after the relevant permutation $(1\ 2\ 3) \mapsto (2 \ 3\ 1)$. Thus, we ask if we can have 
\be\label{3-space-3}
[\lambda e_0 + \mu e_1, e_2, e_3]= [ e_0 , e_2, e_3 + \iC\epsilon \,  e_1] \quad (\epsilon = \pm 1),
\ee
i.e., we ask if, for any $\alpha ,\beta, \gamma  \in \mathbb{C}$, we can find $\alpha' ,\beta',\gamma ' \in \mathbb{C}$ such that
\be\label{3-space-4}
\alpha(\lambda e_0+\mu e_1) +\beta e_2 + \gamma e_3 = \alpha' e_0+ \beta ' e_2 +\gamma' (e_3 + \iC\epsilon\,  e_1),
\ee
or
\be\label{3-space-5}
\alpha \lambda = \alpha ', \quad \alpha \mu = \iC \epsilon \gamma ', \quad \beta =\beta  ', \quad \gamma =\gamma  '.
\ee
Thus we must have $\gamma '=\gamma =-\iC \epsilon \,\alpha \mu $, so once more the relevant equality, here (\ref{3-space-4}), cannot hold on a whole 3-space.\\

3) The remaining possibility is the equality 
\be\label{3-space-6}
[\lambda e_0 + \mu e_2, e_3, e_1]= [ e_0 , e_1, e_2 + \iC\epsilon \,  e_3] \quad (\epsilon = \pm 1).
\ee
The same trivial discussion as for the case (\ref{3-space-3}) leads here (with the same notations) to $\beta  =\iC \epsilon \,\alpha \mu $, so (\ref{3-space-6}) cannot happen. \\

Thus there is just one 3-space that is invariant under both $L_1$ and $L_2$, namely the ``spatial 3-space" $[e_1, e_2, e_3]$. (That 3-space is invariant under $L_3$ as well.) But it clearly is not invariant under the boost $L'_1$, Eq. (\ref{L'_1}). We conclude that no proper subspace of $\mathbb{C}^4$ is invariant under the set $\mathrm{S}$, which is therefore an irreducible set of matrices. This proves \hyperref[Proposition 2]{Proposition 2}. \hfill $\square$\\

\noi It is clear that e.g. the set $\mathrm{S}':=\{L_2, L_3, L'_2\}$, or the set $\mathrm{S}'':= \{L_3,L_1, L'_3\}$ (with $L'_i$ the boost in the direction $e_i$), are irreducible also.

\section{Proof of Lemma 2}\label{ProofLemma2}

Let the endomorphism $T$ of $\mathrm{E}=\mathrm{F}\oplus \mathrm{G}$ be as in the statement of \hyperref[Lemma 2]{Lemma 2}, and let $\mathrm{W}$ be a vector subspace of $\mathrm{E}$ that is invariant under $T$. Suppose first that $\mathrm{W}\subset \mathrm{G}$. Then the fact that $\mathrm{W}$ has indeed the form (\ref{u_i&v'_k}) claimed by \hyperref[Lemma 2]{Lemma 2} is a direct application of the following known result (e.g. \cite{Gohberg2006}):

\paragraph{Lemma 3.} {\it Assume the endomorphism $T$ of the vector space $\mathrm{G}$ admits a basis of eigenvectors   $(v_j)\ (j=1,...,n)$ corresponding with pairwise distinct eigenvalues. Then any subspace of $\mathrm{G}$ that is invariant under $T$ has the form
\be
\mathrm{W} = [(v_j)_{j\in \mathrm{J}}] := \mathrm{Span}\{v_j ; j\in \mathrm{J}\},
\ee
where $\mathrm{J}$ is some subset of $\{1,...,n\}$.}\\

\noi If instead $\mathrm{W}\not\subset \mathrm{G}$, define $\mathrm{W}'$ and $\mathrm{W}''$ as the projection space of $\mathrm{W}$ onto $\mathrm{F}$ or $\mathrm{G}$, respectively, in the decomposition $\mathrm{E}=\mathrm{F}\oplus \mathrm{G}$. This definition can be written explicitly as:
\be\label{W'}
\mathrm{W}' = \{y\in \mathrm{F};\ \exists z \in \mathrm{G}: y+z \in \mathrm{W} \},
\ee
and the like for $\mathrm{W}''$. The subspace $\mathrm{W}'$ is not reduced to zero, for otherwise we would have $\mathrm{W}\subset \mathrm{G}$. Indeed, $\mathrm{E}=\mathrm{F}\oplus \mathrm{G}$ implies that for any $x\in \mathrm{W}$ there is a unique pair $y\in \mathrm{F}$, $z\in \mathrm{G}$ such that $x=y+z$; if $\mathrm{W}' = \{0\}$ we must have $y=0$ by (\ref{W'}). So let $(u_i)_{i=1,...,p}$ ($1\leq p\leq \mathrm{dim\,F}$) be a basis of $\mathrm{W}'$. Since each of the $u_i$ 's belongs to $\mathrm{W}'$, by the definition (\ref{W'}) there is for each of them a vector $z \in \mathrm{G}$ such that the vector $x:=u_i+z$ belongs to $\mathrm{W}$. Let us decompose $z$ on the basis $(v_j)$ of $\mathrm{G}$, thus getting numbers $z_j$  such that $z = \sum _{j=1} ^n z_j\, v_j$. Thus we have
\be\label{x_on_u_i_&_v_j}
x= u_i + \sum _{j=1} ^n v'_j \in \mathrm{W},
\ee
where, by the assumption of \hyperref[Lemma 2]{Lemma 2}, $v'_j := z_j\, v_j$ is still an eigenvector of $T$ for the eigenvalue $\lambda _j$ (even though possibly $v'_j=0$), i.e., $T\,v'_j = \lambda _j\, v'_j$, while $u_i$ is an eigenvector of $T$ for the eigenvalue $\lambda \ne \lambda _j\ (j=1,...,n)$. Thus by \hyperref[Lemma 4]{Lemma 4} below, we have $u_i \in \mathrm{W}$.

\paragraph{Lemma 4\  (e.g. \cite{Stack2015}).}\label{Lemma 4}  {\it Assume that $v_1, ...,v_m$ are eigenvectors of the endomorphism $T$ of the vector space $\mathrm{E}$ corresponding with pairwise distinct eigenvalues $\lambda _j$. If $\mathrm{W}$ is an invariant subspace of $\mathrm{E}$ under $T$ such that $v_1+...+v_m \in \mathrm{W}$, then for each $j =1,...,m$ we have $v_j \in \mathrm{W}$.}\\

\vspace{2mm}
\noi {\it End of the proof of \hyperref[Lemma 2]{Lemma 2}.} Suppose first that 
\be\label{W''}
\mathrm{W}'' := \{z\in \mathrm{G};\ \exists y \in \mathrm{F}: y+z \in \mathrm{W} \}
\ee
is reduced to zero. Then, because $\mathrm{E}=\mathrm{F}\oplus \mathrm{G}$, we have $\mathrm{W}\subset \mathrm{F}$, just like, as we showed after (\ref{W'}), we have $\mathrm{W}\subset \mathrm{G}$ in the symmetric case $\mathrm{W}' = \{0\}$.  Therefore, it is immediate to check that $\mathrm{W}=\mathrm{W}'$, so $(u_i)_{i=1,...,p}$ is a basis of $\mathrm{W}$, hence $\mathrm{W}$ has indeed the form (\ref{u_i&v'_k}). \\

If instead $\mathrm{W}'' \ne \{0\}$, we build a basis $(v_j)_{j\in \mathrm{J}}$ of $\mathrm{W}''$ extracted from the basis $(v_j)_{j=1,...,n}$ of $\mathrm{G}$ (made of eigenvectors of $T$), as follows. For any $x \in \mathrm{W}$, there is a unique pair $y(x)=P_\mathrm{F}(x)\in \mathrm{F}$, $z(x)=P_\mathrm{G}(x)\in \mathrm{G}$ such that 
\be\label{Decomp_x}
x=y(x)+z(x). 
\ee
Since $(v_j)$ is a basis of $\mathrm{G}$, we can decompose $z(x)$ uniquely on the basis $(v_j)_{j=1,...,n}$, so
\be\label{Decomp_z}
z(x) = \sum_{j\in \mathrm{J}(x)} z_j(x)\,v_{j},\quad z_j(x)\ne 0\ \mathrm{for}\ j\in \mathrm{J}(x).
\ee
(Note that $\mathrm{J}(x)$ can be empty, which occurs iff $z(x)=0$.) By a very similar argument, also using \hyperref[Lemma 4]{Lemma 4}, to that developed around Eq. (\ref{x_on_u_i_&_v_j}), we see that (\ref{Decomp_x}) and (\ref{Decomp_z}) imply that $v_j \in \mathrm{W}$ if $j\in \mathrm{J}(x)$. Hence $v_j \in \mathrm{W}''$ if $j\in \mathrm{J}(x)$, since $\mathrm{W}\cap \mathrm{G} \subset \mathrm{W}''$ from (\ref{W''}). Then define
\be
\mathrm{J} := \bigcup _{x\in \mathrm{W} } \mathrm{J}(x).
\ee
Note that we have by construction $\mathrm{J}\subset \{1,...n\}$. We claim that the finite family $(v_j)_{j\in \mathrm{J}}$ is a basis of $\mathrm{W}''$. Indeed, consider any $z''\in \mathrm{W}''$. From (\ref{W''}), there is some $y''\in \mathrm{F}$ such that 
\be
x:= y''+z'' \in \mathrm{W}.
\ee
Hence we have $y''=y(x)$ and $z''=z(x)$ from the uniqueness of the decomposition (\ref{Decomp_x}), so that from (\ref{Decomp_z}):
\be
z'' = \sum_{j\in \mathrm{J}(x)} z_j(x)\,v_{j}.
\ee
Thus the family $(v_j)_{j\in \mathrm{J}}$ does generate $\mathrm{W}''$, and being a free family as an extracted family from the basis $(v_j)_{j=1,...,n}$, it is indeed a basis of $\mathrm{W}''$. Since we showed that $u_i \in \mathrm{W}\ (i=1,...,p)$ and that $v_j \in \mathrm{W}\ (j\in \mathrm{J})$, it is clear that 
\be\label{u_i&v'_k-2}
\mathrm{W} \supset  \mathrm{Span}\{(u_i)_{i=1,...,p};(v_j)_{j\in \mathrm{J}}\}.
\ee
Conversely, note that in the decomposition (\ref{Decomp_x}) of any $x\in \mathrm{W}$, we have $y(x)\in \mathrm{W}'$ from (\ref{W'}) and $z(x)\in \mathrm{W}''$ from (\ref{W''}). Hence the reverse inclusion follows from the fact that $(u_i)_{i\in \{1,...,p\}}$ is a basis of $\mathrm{W}'$ and that $(v_j)_{j\in \mathrm{J}}$ is a basis of $\mathrm{W}''$. This completes the proof of \hyperref[Lemma 2]{Lemma 2}. \hfill $\square$\\

%%%%%%%%%%%%%%%%%%%%%%%%%%%%%%%%%%%%%%%%%%%%%%%%%%%%%%%%%%%%%%%%%%%%%%%%%%%%%%%%

%%%%%%%%%%%%%%%%%%%%%%%%%%%%%%%%%%%%%%%%%%%%%%%%%%%%%%%%%%%%%%%%%%%%%%%%%%%%%%%%
\end{document}